# Charge-spin conversion signal in WTe$_2$ van der Waals hybrid devices with a geometrical design

Bing Zhao[1], Anamul Md. Hoque[1], Dmitrii Khokhriakov[1], Bogdan Karpiak[1], Saroj P. Dash[1,*]

[1]Department of Microtechnology and Nanoscience,
Chalmers University of Technology, SE-41296, Göteborg, Sweden

The efficient generation and control of spin polarization via charge-spin conversion in topological semimetals are desirable for future spintronic and quantum technologies. Here, we report the charge-spin conversion (CSC) signals measured in a Weyl semimetal candidate WTe$_2$ based hybrid graphene device with a geometrical design. Notably, the geometrical angle of WTe$_2$ on the graphene spin-valve channel yields contributions to symmetric and anti-symmetric CSC signal components. The spin precession measurements of CSC signal at different gate voltages and ferromagnet magnetization shows the robustness of the CSC in WTe$_2$ at room temperature. These results can be useful for the design of heterostructure devices and in the architectures of two-dimensional spintronic circuits.

**Keywords:** Weyl-semimetal, Type-II, WTe$_2$, Graphene, van der Waals heterostructures, charge-spin conversion, Geometry effects, nonlocal spin-valve, spin precession.

Corresponding author: **Saroj P. Dash**, Email: saroj.dash@chalmers.se

The spin-orbit interaction (SOI) is a crucial phenomenon in condensed matter physics[1,2] and practical applications in spintronic memory, logic and neuromorphic computing[3–6]. The influence of SOI on the electronic wave functions gives rise to unique topologically protected states[7,8], strong spin-valley coupling[9–12], and spin textures[13]. The spin-dependent electronic scattering from heavy elements and the presence of spin-texture of the electronic bands in the material with large SOI is the origin of the spin Hall effect (SHE)[14,15] and Rashba-Edelstein effect (REE)[16,17]. This charge-spin conversion (CSC) effects due to SHE and REE provide the accumulation of the spin-polarized elections at the boundaries of the solids by application of charge current and have shown potential in spintronics technologies[18].

The topological quantum materials and transition metal dichalcogenides (TMDs) have revealed opportunities to realize several extraordinary CSC phenomena due to their lower crystal symmetry and spin topologies in the band structure[19]. The CSC efficiencies in such materials are recently evaluated using spin-orbit torque (SOT)[20,21], 2nd harmonic Hall[22], and spin potentiometric measurements in local[23–25], and nonlocal geometries[26]. Graphene has been proven to be an excellent candidate for spin transport[27–29]; while TMDs are offering a platform for the efficient CSC and manipulation of spin and valley degrees of freedom[9,12,30]. Therefore, graphene/TMDs heterostructures are popularly adopted to measure the CSC effects in TMDs[31–33]. Recent experiments also show that the TMDs can imprint proximity effects on graphene, like SOI[34–36], magnetism[37,38], spin lifetime anisotropy[12,30], and CSC process due to SHE and REE[39–44].

Recently, theoretical prediction[45–47] and experimental observation of both conventional[33,39–42,48,49] and unconventional[50–52] CSC effects have been measured in semimetals of $WTe_2$ and $MoTe_2$. For the detection of CSC effects in $WTe_2$, graphene-based hybrid spin-valve devices were used in the spin precession Hanle geometry[53]. A vertical geometry is also introduced for the detection of CSC phenomenon in TMDs [50,54,55]. The experiments on $WTe_2$ demonstrated the detection of unconventional CSC and efficient spin injection and detection in the graphene channel[50]. For device applications, such CSC based spin-polarized $WTe_2$ sources/detectors should be integrated with two-dimensional (2D) device architectures in different geometries.

Here, we report the evolution of conventional CSC signal components in a Weyl semimetal $WTe_2$ device with a geometrical design. Using a $WTe_2$/graphene heterostructure spin-valve device, we made a rotation angle of the $WTe_2$ spin detector relative to the ferromagnetic injector and graphene channel. We investigated the detailed spin precession signal for the detection of conventional inverse CSC (ICSC) in $WTe_2$ with different gate voltages and initial magnetization orientations of the ferromagnetic spin injector. The observation of symmetric and anti-symmetric ICSC signal components due to a geometric angle of $WTe_2$ in the device design provides a perspective for better understanding and its utilization in an all-electrical van der Waals spintronic devices.

Figures 1a and 1b show the hybrid device, where $WTe_2$ was placed at a tilt angle φ on the graphene channel. The $WTe_2$ single-crystal flakes are in the $T_d$-bulk phase at room temperature. Graphene channel was obtained by the pattering of CVD graphene, followed by an $O_2$ plasma etching process. The thin $WTe_2$ flakes were then transferred onto the graphene strips by the dry transfer method inside a glovebox in a controlled environment. Ferromagnetic electrodes (FM) Co/$TiO_2$ (contacts 2,3,6) and non-magnetic contacts Au/Ti (contacts 1,4,5,7) were prepared by electron beam lithography and e-beam evaporation, followed by lift-off methods (see Supplementary information). Ferromagnetic tunnel contacts resistances on graphene were in the range of 2-10 kΩ, and the WTe$_2$/graphene interface resistance was ~25Ω at room temperature.

First, we measured the spin injection, transport, and detection in the pristine graphene channel using nonlocal spin valve geometry. The bias current was applied between contact 7 and 3, a nonlocal voltage signal was detected via contacts 2 and 1 (labeled as NL-7321, see Fig. 1c). As expected, a clear spin-valve signal ($R_{nl}$~0.18Ω) was present in the graphene channel. However, no spin valve signal was observed for measurement across the WTe$_2$/graphene heterostructure (labeled as NL-1367), indicating an apparent spin absorption by WTe$_2$ at the heterostructure region[56–58] (Fig.1c).

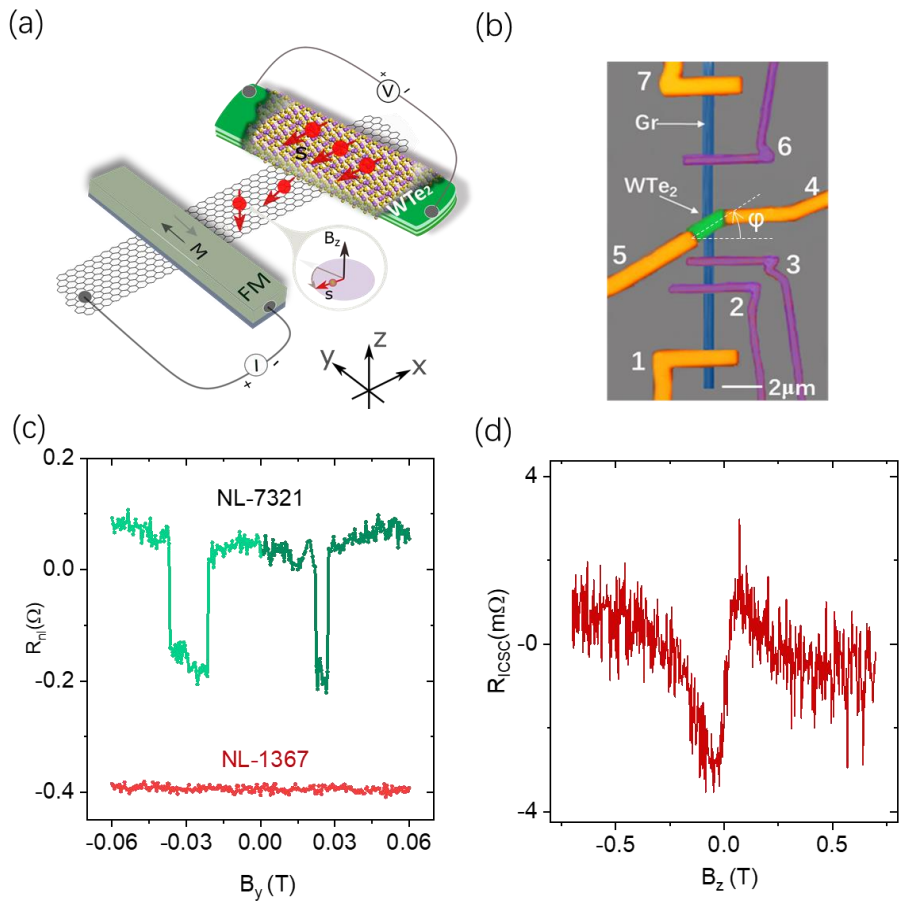

*Figure 1. A Geometrical design of charge-spin conversion signal in WTe$_2$. (a) Schematic of the inverse charge-spin conversion (ICSC) measurement geometry, with a tilt angle of the WTe$_2$ crystal on graphene channel, the spin injector ferromagnet, spin **s** orientations. The inset shows the spin precession driven by the out-of-plane magnetic field $B_z$. (b) Colored optical microscope picture of the nanofabricated device for electrical detection of the nonlocal spin-valve and spin-charge conversion in WTe$_2$. The WTe$_2$ crystal is placed at an angle φ on the graphene channel with ferromagnetic (TiO$_2$/Co, purple) and non-ferromagnetic (Ti/Au, golden) contacts. (c) The nonlocal spin-valve signal in the graphene channel section with contacts 2,3 with $R_{nl}=V_{21}/I_{73}$ and section with contacts 6,3, with $R_{nl}=V_{67}/I_{13}$, respectively. (d) The ICSC spin precession signals ($R_{ICSC}=V_{54}/I_{31}$) as measured using geometry shown in schematics (a) with injection bias current I=100 µA at gate voltage $V_g$=-60V. All the measurements are performed at room temperature.*

The inverse charge-to-spin conversion (ICSC) measurements were performed at room temperature using a geometry shown in Fig. 1a, where the spin current is injected from ferromagnet (contact 3) into the graphene channel. The spin current diffuse towards the $WTe_2$/graphene heterostructure region and gets absorbed by the $WTe_2$ flake. Consequently, the spin current in $WTe_2$ gives rise to a charge voltage due to spin-to-charge conversion (ICSC) as detected between the two ends (contacts 4 and 5) of $WTe_2$. The ICSC voltage signal was measured as a function of out-of-plane magnetic field $B_z$, where the spins in the graphene channel precess before arriving at the $WTe_2$/graphene junction. This spin precession induces the evolution of the signal $R_{ICSC}=V_{54}/I_{31}$ in a sine-shaped curve with the magnetic field $B_z$, as shown in Fig. 1d. Consistent with our previous measurements, a conventional ICSC signal is observed in the condition of very low $WTe_2$/graphene interface resistance ~ 25Ω. We notice that the ICSC data is not entirely anti-symmetric about $B_z=0T$, which is due to an angle φ existing between $WTe_2$ flake and graphene channel and as well as with the ferromagnetic contacts. The detailed gate dependence of the out-of-plane ICSC signal $R_{ICSC}$ was measured, as shown in Fig. 2a. As expected, the $R_{ICSC}$ signal does not change the sign for electron and hole-doped graphene regime, again confirming the spin origin of the ICSC signal, as reported in our previous work[30]. As $WTe_2$ is semi-metallic (i.e., not sensitive to the gate voltage), the observed gate dependent variation of $\Delta R_{ICSC}$ (extracted from Eq. 2) can be mainly due to modulation graphene channel resistance and conductivity mismatch issues with FM tunnel contacts on graphene (Fig. 2b)[59]. According to the spin injection and detection principle, the impedance mismatch[59,60] between the ferromagnet and graphene causes a change in the spin injection efficiency. The smaller spin signal in the positive gate voltage regime and higher graphene resistance affects the noise level of the measured ICSC signal.

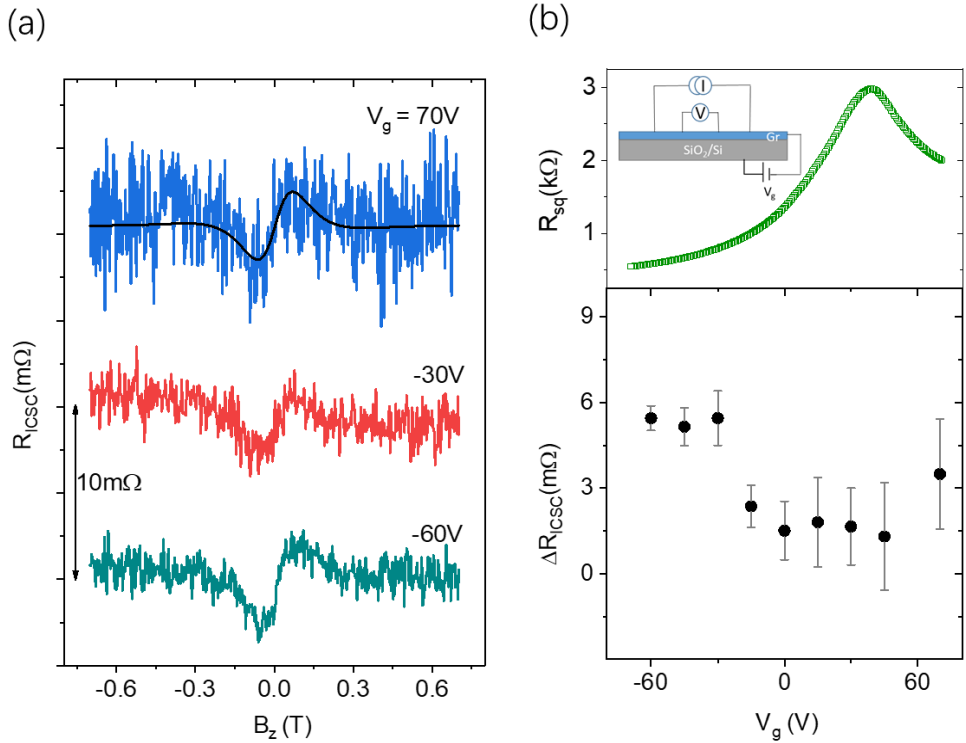

*Figure 2. Back gate dependence of the charge-spin conversion signal. (a) The measured ICSC signal $R_{ICSC}$ for different gate voltages at room temperature, with a shift in the y-axis for the sake of clarity. The black line in $V_g$ of 70V data is a guide to the eye. (b) Gate dependence of the graphene channel resistance $R_{sq}$ and the ICSC signal magnitude $\Delta R_{ICSC}$. The inset in the upper panel is the measurement geometry of the gate*

*dependence of the graphene channel.*

As mentioned, we purposefully designed the device with a tilted WTe$_2$ flake (Fig. 3a), to investigate the evolution of geometry induced (I)CSC signal components. The measured ICSC curve in Fig. 3b is not strictly anti-symmetric about B$_z$=0T. This is due to the initialization of the direction of the spins relative to the WTe$_2$. In a standard geometry, the spins are parallel to the WTe$_2$ at B$_z$=0T; while here, an angle φ exists, causing the phase shift of the spin precession relative to the usual parallel geometry. Figure 3b also shows the magnetization orientation (±M) dependence of the ICSC signal, confirming the spin origin of the measured signal. The initialization of ferromagnetic contact magnetization direction (±M) defines the shape of the sine-like curve with opposite directions. To remove any signal which does not come from the y-component of the injector magnetization (±M), we obtain the averaged signal by R$_{ICSC, averaged}$=(R$_{ICS,+M}$-R$_{ICSC,-M}$)/2 (see Fig. 3c). Mathematically, we further decompose the signal and obtain the symmetric (sym) and anti-symmetric (asym) component (Fig. 3c) by using the formula below[61,62],

$$R_{ICSC}(sym) = [R_{ICSC}(B) + R_{ICSC}(-B)]/2; \quad R_{ICSC}(asym) = [R_{ICSC}(B) - R_{ICSC}(-B)]/2. \quad \text{Eq. (1)}$$

We obtain the tilted angle of WTe$_2$ to be φ= arctan(R$_{sym}$/R$_{asym}$)=25°±10°, where R$_{sym}$ and R$_{asym}$ are the magnitude of the symmetric and anti-symmetric component signals, respectively (R$_{asym}$ is defined as the difference of the extrema). This experimental estimation of the geometry of the WTe$_2$ flake in the device agrees well with the optical image result (~30°, see Fig.1b). To be noted, our model is based on the 1D assumption of the WTe$_2$ flake, i.e. the width can be ignored compared to its length.

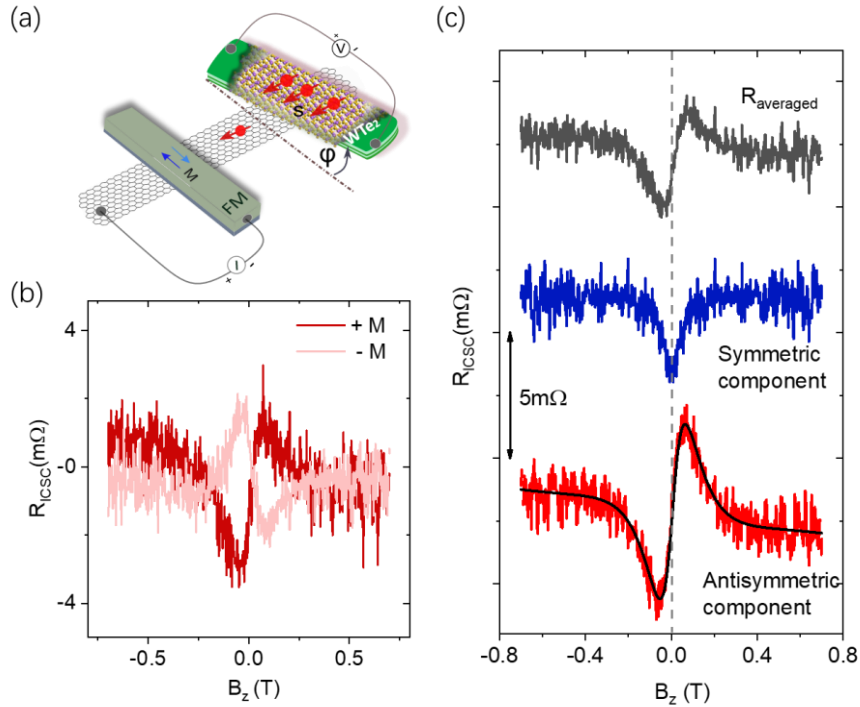

**Figure 3. Effect of geometry on the detection of the ICSC signal in the WTe$_2$.** *(a) Schematics of ICSC measurement geometry with spin precession in the graphene channel and the effective tilted angle φ of WTe$_2$ relative to FM electrode. (b) The effect of magnetic moment direction (+/-M) on the spin precession-induced ICSC signal at gate voltage V$_g$=-60V. (c) The averaged ICSC signal R$_{ICSC, averaged}$=(R$_{-M}$-R$_{+M}$)/2, the extracted*

*symmetric (blue) and anti-symmetric (red) components of the signal. The black curve is the fitting result with Eq. (2).*

To quantitatively analyze the ICSC signal, we adopt the out-of-plane ICSC Hanle precession formula to fit the anti-symmetric component[56],

$$R_{ICSC}(B_z) = \Delta R_{ICSC} \lambda_{WTe2} \left[\frac{1-\exp(-\frac{t_M}{\lambda_{WTe2}})}{1+\exp(-\frac{t_M}{\lambda_{WTe2}})}\right] \int_0^\infty \frac{\exp(-\frac{L_{SH}^2}{4D_s t})}{t\sqrt{4\pi D_s t}} \sin(\omega_0 B_z t) \exp(-\frac{t}{\tau_s}) dt,$$

$$= \frac{p_{eff} \theta_{ICSC} \rho_M L_{SH} \lambda_{WTe2}}{w_M t_M} \left[\frac{1-\exp(-\frac{t_M}{\lambda_{WTe2}})}{1+\exp(-\frac{t_M}{\lambda_{WTe2}})}\right] \int_0^\infty \frac{\exp(-\frac{L_{SH}^2}{4D_s t})}{t\sqrt{4\pi D_s t}} \sin(\omega_0 B_z t) \exp(-\frac{t}{\tau_s}) dt$$

Eq. (2)

Here we define the ICSC magnitude $\Delta R_{ICSC} = \frac{p_{eff} \theta_{ICSC} \rho_M L_{SH}}{w_M t_M}$, Where $\rho_M$, $w_M$, $t_M$, $\theta_{SH}$ and $\lambda_{WTe2}$ are resistivity, width, thickness, spin Hall angle, and spin diffusion length of WTe$_2$, respectively. $L_{SH}$ is graphene channel length, $p_{eff}$=0.057 and $D_s$=0.030m$^2$/s are the effective spin polarization of Co/TiO$_2$ and spin diffusion constant, extracted from the standard fitting of the Hanle signal in graphene spin valve nearby the WTe$_2$ (see Supplementary Information Fig. S1). Here, we D$_s$ is the spin diffusion constant, $\tau_s$ is the spin diffusion time in graphene, $\omega_0 = g\mu_B/\hbar$ is the Larmor precession frequency, where g=2 is Landé g-factor, $\mu_B$ and $\hbar$ are the Bohr magneton and reduced Planck constant. To be noted, we have the relation $R_{ICSC}(\text{asym}) = \cos(\varphi) \cdot R_{ICSC}(B_z)$ in the fitting process with φ=30°. By fitting the anti-symmetric component, we obtain the Edelstein length $\lambda_{EE}=\theta_{ICSC}\lambda_{WTe2} \approx 1.016 \pm 0.004$ nm, which agrees well with previously reported values of WTe$_2$[33].

To systematically investigate the WTe$_2$ tilted angle dependence of the signal, we have mathematically calculated the results via the projection of the effective spins on the WTe$_2$ flake. This is based on a simple analytical expression[61,62], R$_{ICSC}$ ∝ cos(θ), where the angle θ is defined as the angle between the initial spins and the detector (WTe$_2$). Now a titled angle φ of the detector WTe$_2$ is considered. Mathematically, one can decompose the detected signal into symmetric (averaged spins // WTe$_2$) R$_{ICSC}$(sym) and asymmetric (averaged spins ⊥ WTe$_2$) components R$_{ICSC}$(asym). In the calculation, the symmetric component was taken from a normalized Hanle curve; while the anti-symmetric component from a normalized asymmetric Hanle curve[61,62]. The calculated results show that the symmetric and anti-symmetric components have the cosine and sine relation with the tilted angle φ, respectively (Fig. 4b). Consequently, we can obtain the resultant R$_{ICSC}$ by R$_{ICSC}$=(R$_{ICSC}$(sym)+ R$_{ICSC}$(asym))/2. The calculated resultant ICSC signals can be tuned persistently with the evolution from a sine shape Hanle curve to a cosine behavior with positive titled angles φ (anticlockwise rotation) (Fig. 4c). In contrast, a reversed sign of the Hanle curve is observed in the negative angles -φ (clockwise rotation) of the WTe$_2$ flake (Fig. 4d). This calculation shows that the ICSC signal is expected to be evolved like in figures 4(b-d) as a function of the angle of WTe$_2$ flake, and further experimental studies with multiple angles are required. Such developments will have a great potential to tune spin signal via the use of the geometry of the WTe$_2$ on a graphene channel.

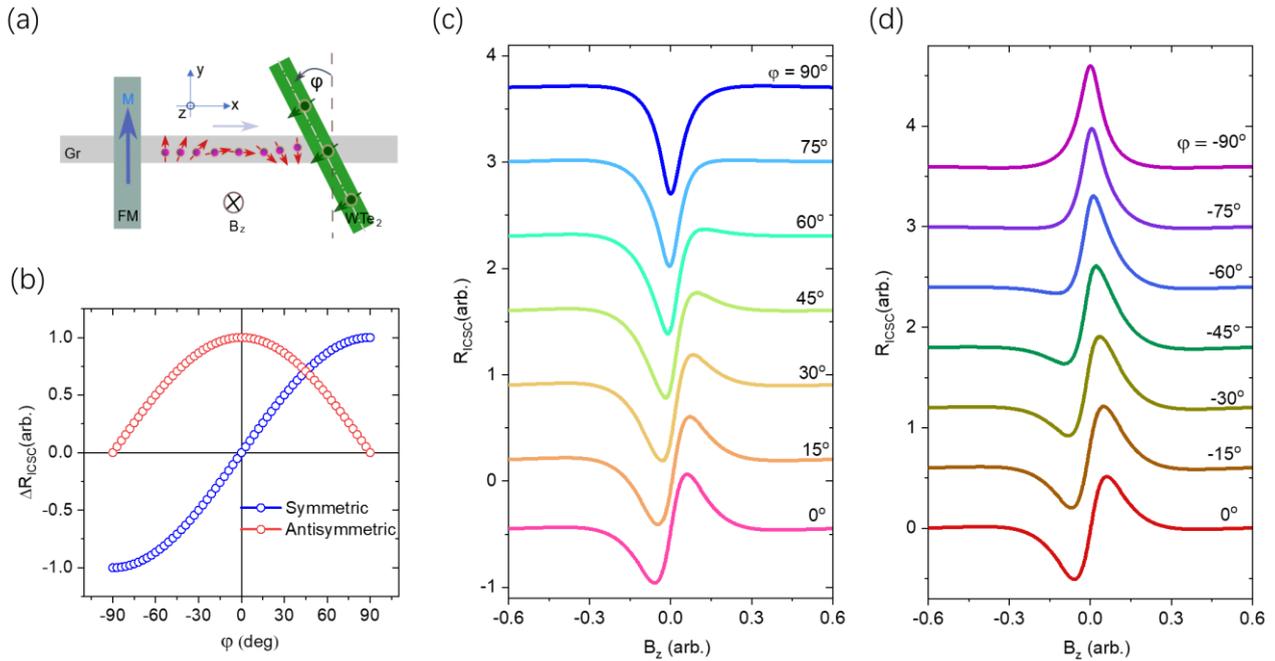

*Figure 4. Simulation of the evolution of the ICSC signal with the tilted angle of WTe$_2$ crystal on a graphene spin-valve. (a) The top view of the ICSC measurement geometry with tilted WTe$_2$ crystal at an angle φ, magnetic moment M of the FM spin injector, spin precession in graphene channel, and diffusion towards the WTe$_2$/Gr heterostructure. (b) The calculated tilted angle φ dependence of the normalized symmetric and anti-symmetric components. (c, d) The calculated ICSC signals with different tilted angle φ (+90 to -90 degree) of WTe$_2$ crystal.*

In conclusion, we demonstrated robust charge-spin conversion in WTe$_2$ with nonlocal spin precession measurements, supported by the detailed gate voltage and injector magnetization orientation investigations. By introducing a geometrical angle of the WTe$_2$ in the graphene spin-valve devices, we could reveal the effect of the geometry on the observation of symmetric and anti-symmetric charge-spin conversion signal components. Our calculations show that the signal magnitudes can be tuned by controlling the geometrical angles of the spin-source or -detector. With the recent realization of graphene spin interconnect[27] and proposed spin logic technologies[5,63], the realization of the nonmagnetic spin injector/detector with WTe$_2$[33,50] and its geometrical design can provide substantial advances in the spin-based nanoelectronics device concepts. These results provide a perspective for the spintronic device design for injection and detection of spin polarization for read/write functionalities in 2D circuit architectures.

**Data Availability Statement**
The data that support the findings of this study are available from the corresponding author upon reasonable request.

**Supplementary Material**
See Supplementary Material for the device fabrication, measurement methods and Hanle measurement on graphene channel.

**Acknowledgments**

The authors acknowledge financial supports from EU Graphene Flagship (Core 2 No. 785219, and Core 3 No. 881603), Swedish Research Council VR project grants (No. 2016-03658), 2D Tech VINNOVA center, Graphene center, and the EI Nano and AoA Materials program at Chalmers University of Technology.

**References**

[1] A. Soumyanarayanan, N. Reyren, A. Fert, and C. Panagopoulos, Nature **539**, 509 (2016).

[2] A. Manchon, H.C. Koo, J. Nitta, S.M. Frolov, and R.A. Duine, Nat Mater **14**, 871 (2015).

[3] A. Manchon, J. Železný, I.M. Miron, T. Jungwirth, J. Sinova, A. Thiaville, K. Garello, and P. Gambardella, Rev Mod Phys **91**, 035004 (2019).

[4] J. Grollier, D. Querlioz, K.Y. Camsari, K. Everschor-Sitte, S. Fukami, and M.D. Stiles, Nat Electron **3**, 360 (2020).

[5] B. Behin-Aein, D. Datta, S. Salahuddin, and S. Datta, Nat Nanotechnol **5**, 266 (2010).

[6] S. Manipatruni, D.E. Nikonov, C.-C. Lin, T.A. Gosavi, H. Liu, B. Prasad, Y.-L. Huang, E. Bonturim, R. Ramesh, and I.A. Young, Nature **565**, 35 (2019).

[7] M.Z. Hasan and C.L. Kane, Rev Mod Phys **82**, 3045 (2010).

[8] B. Yan and S.-C. Zhang, Reports Prog Phys **75**, 096501 (2012).

[9] A.W. Cummings, J.H. Garcia, J. Fabian, and S. Roche, Phys Rev Lett **119**, 206601 (2017).

[10] M. Gmitra and J. Fabian, Phys Rev Lett **119**, 146401 (2017).

[11] T.S. Ghiasi, J. Ingla-Aynés, A.A. Kaverzin, and B.J. Van Wees, Nano Lett **17**, 7528 (2017).

[12] L.A. Benítez, J.F. Sierra, W. Savero Torres, A. Arrighi, F. Bonell, M. V. Costache, and S.O. Valenzuela, Nat Phys **14**, 303 (2018).

[13] H. Maaß, H. Bentmann, C. Seibel, C. Tusche, S. V. Eremeev, T.R.F. Peixoto, O.E. Tereshchenko, K.A. Kokh, E. V. Chulkov, J. Kirschner, and F. Reinert, Nat Commun **7**, 11621 (2016).

[14] J. Sinova, S.O. Valenzuela, J. Wunderlich, C.H. Back, and T. Jungwirth, Rev Mod Phys **87**, 1213 (2015).

[15] T. Jungwirth, J. Wunderlich, and K. Olejník, Nat Mater **11**, 382 (2012).

[16] M. Isasa, M.C. Martínez-Velarte, E. Villamor, C. Magén, L. Morellón, J.M. De Teresa, M.R. Ibarra, G. Vignale, E. V. Chulkov, E.E. Krasovskii, L.E. Hueso, and F. Casanova, Phys Rev B **93**, 23 (2016).

[17] J.C.R. Sánchez, L. Vila, G. Desfonds, S. Gambarelli, J.P. Attané, J.M. De Teresa, C. Magén, and A. Fert, Nat Commun **4**, 2944 (2013).

[18] W. Han, Y. Otani, and S. Maekawa, Npj Quantum Mater **3**, 27 (2018).

[19] B. Yan and C. Felser, Annu Rev Condens Matter Phys **8**, 337 (2017).

[20] H. Wu, P. Zhang, P. Deng, Q. Lan, Q. Pan, S.A. Razavi, X. Che, L. Huang, B. Dai, K. Wong, X. Han, and K.L. Wang, Phys Rev Lett **123**, 207205 (2019).

[21] D. MacNeill, G.M. Stiehl, M.H.D. Guimaraes, R.A. Buhrman, J. Park, and D.C. Ralph, Nat Phys **13**, 300 (2017).

[22] Q. Shao, G. Yu, Y.-W. Lan, Y. Shi, M.-Y. Li, C. Zheng, X. Zhu, L.-J. Li, P.K. Amiri, and K.L. Wang, Nano Lett **16**, 7514 (2016).

[23] A. Dankert, P. Bhaskar, D. Khokhriakov, I.H. Rodrigues, B. Karpiak, M.V. Kamalakar, S. Charpentier, I. Garate, and S.P. Dash, Phys Rev B **97**, 125414 (2018).

[24] A. Dankert, J. Geurs, M.V. Kamalakar, S. Charpentier, and S.P. Dash, Nano Lett **15**, 7976 (2015).

[25] P. Li, W. Wu, Y. Wen, C. Zhang, J. Zhang, S. Zhang, Z. Yu, S.A. Yang, A. Manchon, and X. Zhang, Nat Commun **9**, 3990 (2018).

[26] K. Vaklinova, A. Hoyer, M. Burghard, and K. Kern, Nano Lett **16**, 2595 (2016).

[27] D. Khokhriakov, B. Karpiak, A.M. Hoque, B. Zhao, S. Parui, and S.P. Dash, ACS nano http://doi.org/10.1021/acsnano.0c07163 (2020).

[28] M.V. Kamalakar, C. Groenveld, A. Dankert, and S.P. Dash, Nat Commun **6**, 6766 (2015).

[29] Z.M. Gebeyehu, S. Parui, J.F. Sierra, M. Timmermans, M.J. Esplandiu, S. Brems, C. Huyghebaert, K. Garello, M. V. Costache, and S.O. Valenzuela, 2D Mater **6**, 034003 (2019).

[30] T.S. Ghiasi, J. Ingla-Aynés, A.A. Kaverzin, and B.J. van Wees, Nano Lett **17**, 7528 (2017).

[31] W. Yan, O. Txoperena, R. Llopis, H. Dery, L.E. Hueso, and F. Casanova, Nat Commun **7**, 13372 (2016).

[32] A. Dankert and S.P. Dash, Nat Commun **8**, 16093 (2017).

[33] B. Zhao, D. Khokhriakov, Y. Zhang, H. Fu, B. Karpiak, A.M. Hoque, X. Xu, Y. Jiang, B. Yan, and S.P. Dash, Phys Rev Res **2**, 013286 (2020).

[34] Z. Wang, D. Ki, H. Chen, H. Berger, A.H. MacDonald, and A.F. Morpurgo, Nat Commun **6**, 8339 (2015).

[35] B. Yang, M.-F. Tu, J. Kim, Y. Wu, H. Wang, J. Alicea, R. Wu, M. Bockrath, and J. Shi, 2D Mater **3**, 031012 (2016).

[36] D. Khokhriakov, A.W. Cummings, K. Song, M. Vila, B. Karpiak, A. Dankert, S. Roche, and S.P. Dash, Sci Adv **4**, eaat9349 (2018).

[37] J.C. Leutenantsmeyer, A.A. Kaverzin, M. Wojtaszek, and B.J. van Wees, 2D Mater **4**, 014001 (2016).

[38] B. Karpiak, A.W. Cummings, K. Zollner, M. Vila, D. Khokhriakov, A.M. Hoque, A. Dankert, P. Svedlindh, J. Fabian, S. Roche, and S.P. Dash, 2D Mater **7**, 015026 (2019).

[39] L.A. Benítez, W. Savero Torres, J.F. Sierra, M. Timmermans, J.H. Garcia, S. Roche, M. V. Costache, and S.O. Valenzuela, Nat Mater **19**, 170 (2020).

[40] A.M. Hoque, D. Khokhriakov, B. Karpiak, and S.P. Dash, Prepr ArXiv190809367v2 (2019).

[41] L. Li, J. Zhang, G. Myeong, W. Shin, H. Lim, B. Kim, S. Kim, T. Jin, S. Cavill, B.S. Kim, C. Kim, J. Lischner,

A. Ferreira, and S. Cho, ACS Nano **14**, 5251 (2020).

[42] D. Khokhriakov, A.M. Hoque, B. Karpiak, and S.P. Dash, Nat Commun **11**, 3657 (2020).

[43] C.K. Safeer, J. Ingla-Aynés, N. Ontoso, F. Herling, W. Yan, L.E. Hueso, and F. Casanova, Nano Lett **20**, 4573 (2020).

[44] F. Herling, C.K. Safeer, J. Ingla-Aynés, N. Ontoso, L.E. Hueso, and F. Casanova, APL Mater **8**, 071103 (2020).

[45] J. Zhou, J. Qiao, A. Bournel, and W. Zhao, Phys Rev B **99**, 060408 (2019).

[46] J.H. Garcia, M. Vila, C.-H. Hsu, X. Waintal, V.M. Pereira, and S. Roche, ArXiv Prepr ArXiv: 2007. 05626 (2020).

[47] M. Vila, C.-H. Hsu, J.H. Garcia, L.A. Benítez, X. Waintal, S. Valenzuela, V.M. Pereira, and S. Roche, ArXiv Prepr ArXiv: 2007.02053 1 (2020).

[48] C.K. Safeer, J. Ingla-Aynés, F. Herling, J.H. Garcia, M. Vila, N. Ontoso, M.R. Calvo, S. Roche, L.E. Hueso, and F. Casanova, Nano Lett **19**, 1074 (2019).

[49] T.S. Ghiasi, A.A. Kaverzin, P.J. Blah, and B.J. van Wees, Nano Lett **19**, 5959 (2019).

[50] B. Zhao, B. Karpiak, D. Khokhriakov, A. Johansson, A.M. Hoque, X. Xu, Y. Jiang, I. Mertig, and S.P. Dash, Adv Mater **2000818**, 2000818 (2020).

[51] C.K. Safeer, N. Ontoso, J. Ingla-Aynés, F. Herling, V.T. Pham, A. Kurzmann, K. Ensslin, A. Chuvilin, I. Robredo, M.G. Vergniory, F. de Juan, L.E. Hueso, M.R. Calvo, and F. Casanova, Nano Lett **19**, 8758 (2019).

[52] P. Song, C.-H. Hsu, G. Vignale, M. Zhao, J. Liu, Y. Deng, W. Fu, Y. Liu, Y. Zhang, H. Lin, V.M. Pereira, and K.P. Loh, Nat Mater **19**, 292 (2020).

[53] C. Wang, Y. Zhang, J. Huang, S. Nie, G. Liu, A. Liang, Y. Zhang, B. Shen, J. Liu, C. Hu, Y. Ding, D. Liu, Y. Hu, S. He, L. Zhao, L. Yu, J. Hu, J. Wei, Z. Mao, Y. Shi, X. Jia, F. Zhang, S. Zhang, F. Yang, Z. Wang, Q. Peng, H. Weng, X. Dai, Z. Fang, Z. Xu, C. Chen, and X.J. Zhou, Phys Rev B **94**, 241119 (2016).

[54] Z. Kovács-Krausz, A.M. Hoque, P. Makk, B. Szentpéteri, M. Kocsis, B. Fülöp, M.V. Yakushev, T.V. Kuznetsova, O.E. Tereshchenko, K.A. Kokh, I.E. Lukács, T. Taniguchi, K. Watanabe, S.P. Dash, and S. Csonka, Nano Lett **20**, 4782 (2020).

[55] A.M. Hoque, D. Khokhriakov, B. Karpiak, and S.P. Dash, Phys Rev Res **2**, 033204 (2020).

[56] W. Savero Torres, J.F. Sierra, L.A. Benítez, F. Bonell, M. V Costache, and S.O. Valenzuela, 2D Mater **4**, 041008 (2017).

[57] Y. Niimi, M. Morota, D.H. Wei, C. Deranlot, M. Basletic, A. Hamzic, A. Fert, and Y. Otani, Phys Rev Lett **106**, 126601 (2011).

[58] W. Yan, E. Sagasta, M. Ribeiro, Y. Niimi, L.E. Hueso, and F. Casanova, Nat Commun **8**, 661 (2017).

[59] W. Han, K. Pi, K.M. Mccreary, Y. Li, J.J.I. Wong, A.G. Swartz, and R.K. Kawakami, Phys Rev Lett **105**, 167202 (2010).

[60] M.H.D.G. and B.J. van W. T. Maassen,I. J. Vera-Marun, Phys Rev B **86**, 235408 (2012).

[61] B. Zhao, D. Khokhriakov, B. Karpiak, A.M. Hoque, L. Xu, L. Shen, Y.P. Feng, X. Xu, Y. Jiang, and S.P. Dash, 2D Mater **6**, 035042 (2019).

[62] S. Ringer, S. Hartl, M. Rosenauer, T. Völkl, M. Kadur, F. Hopperdietzel, D. Weiss, and J. Eroms, Phys Rev B **97**, 205439 (2018).

[63] S. Manipatruni, D.E. Nikonov, C.-C. Lin, T.A. Gosavi, H. Liu, B. Prasad, Y.-L. Huang, E. Bonturim, R. Ramesh, and I.A. Young, Nature **565**, 35 (2019).